# Adversarial Deep Network Embedding for Cross-network Node Classification

**Xiao Shen**
The Hong Kong Polytechnic University
xiao.shen@connect.polyu.hk

**Quanyu Dai**
The Hong Kong Polytechnic University
quanyu.dai@connect.polyu.hk

**Fu-lai Chung**
The Hong Kong Polytechnic University
cskchung@comp.polyu.edu.hk

**Wei Lu**
University of Electronic Science and Technology of China
luwei@uestc.edu.cn

**Kup-Sze Choi**
The Hong Kong Polytechnic University
thomasks.choi@polyu.edu.hk

## Abstract

In this paper, the task of cross-network node classification, which leverages the abundant labeled nodes from a source network to help classify unlabeled nodes in a target network, is studied. The existing domain adaptation algorithms generally fail to model the network structural information, and the current network embedding models mainly focus on single-network applications. Thus, both of them cannot be directly applied to solve the cross-network node classification problem. This motivates us to propose an adversarial cross-network deep network embedding (ACDNE) model to integrate adversarial domain adaptation with deep network embedding so as to learn network-invariant node representations that can also well preserve the network structural information. In ACDNE, the deep network embedding module utilizes two feature extractors to jointly preserve attributed affinity and topological proximities between nodes. In addition, a node classifier is incorporated to make node representations label-discriminative. Moreover, an adversarial domain adaptation technique is employed to make node representations network-invariant. Extensive experimental results demonstrate that the proposed ACDNE model achieves the state-of-the-art performance in cross-network node classification.

## Introduction

Networks, a powerful means to represent complex interactions and relations between entities, are ubiquitous in the real world, such as social networks, citation networks, and protein-protein interaction networks. Cross-network node classification, which transfers the knowledge learned from a source network to help predict node labels in a target network, can benefit a wide variety of applications. For example, in online social networks, given a mature source network with plenty of users having annotated labels indicating their interests, and a newly formed target network short of labels, it would be beneficial to transfer useful knowledge from the source network to make appropriate recommendations to unlabeled users in the target network. In addition, in protein-protein interaction networks, one can leverage the abundant functional information from a source network to help predict the functionalities of proteins in a newly formed target network.

In recent years, domain adaptation has received a lot of attentions. Given a target domain short of labels, domain adaptation aims to leverage the abundant labeled data from a source domain to help target domain learning (Pan and Yang 2010). A popular type of domain adaptation algorithms is feature-based (Long et al. 2013; Long et al. 2015; Ganin et al. 2016; Tzeng et al. 2017; Shen et al. 2018), which aims to learn domain-invariant feature representations to mitigate domain discrepancy. In these domain adaptation algorithms, each data sample is considered as independent and identically distributed during representation learning. This is appropriate for image or text data in computer vision (CV) and natural language processing (NLP). However, in network structural data, each instance (i.e. node) naturally has complicated interactions or relations (i.e. edges) w.r.t. other instances (i.e. its neighbors). It should be rather important and necessary to consider the complex network relationships between nodes for various graph mining tasks. Thus, the existing domain adaptation algorithms which fail to model the network structural information would yield unsatisfactory performance in cross-network node classification.

Recently, network embedding has become an effective method to learn low-dimensional representations which can well preserve the original network structures. Then, one can employ the machine learning algorithms on the low-dimensional embedding representations to solve diverse graph mining applications, such as node classification, link prediction, node clustering and recommendation (Perozzi, Al-Rfou, and Skiena 2014; Cao, Lu, and Xu 2016; Wang, Cui, and Zhu 2016). However, existing network embedding algorithms have been mostly designed for a single-network scenario. When a cross-network scenario is considered, the varied data distributions across networks would pose an obstacle for applying a model learned from a source domain to a target domain (Pan and Yang 2010). Thus, the single-network-embedding algorithms without addressing domain discrepancy would fail to learn transferable representations for cross-network node classification (Shen et al. 2019).

In this work, we address a **cross-network node classification problem**, where given a source network with fully labeled nodes and a target network with completely unlabeled nodes, we aim to learn appropriate node representations based upon which the abundant labeled data from the source network can be successfully leveraged to classify nodes in the target network. To this aim, we propose an adversarial cross-network deep network embedding (ACDNE) model to innovatively integrate deep network embedding with adversarial domain adaptation. The proposed deep network embedding module contains two feature extractors, which learn node representations based on each node's own attributes and its neighbors' attributes weighted by the associated topological proximities respectively. Then, both the attributed affinity and topological proximities between nodes can be well preserved. The same deep network embedding module (i.e. shared trainable parameters) is utilized to generate node representations for the source network and the target network. In addition, a node classifier is incorporated by ACDNE to leverage the supervised signals from the source network to make node representations label-discriminative for node classification. To address the distribution discrepancy across networks, a domain discriminator is incorporated by ACDNE to compete against the deep network embedding module. On one hand, the domain discriminator tries to distinguish the node representations of the source network from those of the target network. On the other hand, the deep network embedding module is trained to learn network-invariant node representations to fool the domain discriminator. Finally, both label-discriminative and network-invariant node representations can be obtained by ACDNE to effectively solve the cross-network node classification problem. The contributions of this work can be summarized as follows:

1) ACDNE is among the first to integrate deep network embedding with adversarial domain adaptation to learn label-discriminative and network-invariant representations for cross-network node classification;

2) The proposed deep network embedding module effectively captures topological proximities and attributed affinity between nodes within a network and across networks;

3) Extensive experimental results in the real-world datasets verify the effectiveness of the proposed ACDNE model for cross-network node classification.

## Related Work

### Domain Adaptation

Early domain adaptation approaches are instance-based, which reweight or subsample instances from the source domain to match the distribution of the target domain (Dai et al. 2007; Huang et al. 2007). Recently, several deep domain adaptation algorithms have been proposed to embed domain adaptation components into deep neural networks to learn domain-invariant representations. They can be categorized as statistic-based and adversarial learning. On one hand, the statistic-based approaches (Long et al. 2013; Long et al. 2015) widely incorporate the Maximum Mean Discrepancy (MMD) metric (Gretton et al. 2007) into deep neural networks to match the mean of the distributions across domains. On the other hand, motivated by the idea of GAN (Goodfellow et al. 2014), the adversarial domain adaptation models (Ganin et al. 2016; Tzeng et al. 2017; Shen et al. 2018) utilize an adversarial loss to minimize the domain shift, where a domain discriminator and a feature extractor compete against each other in a minimax game.

### Network Embedding

Previous network embedding models (Perozzi, Al-Rfou, and Skiena 2014; Tang et al. 2015; Cao, Lu, and Xu 2016; Grover and Leskovec 2016; Wang, Cui, and Zhu 2016; Shen and Chung 2017; Dai et al. 2018; Shen and Chung 2018; Dai et al. 2019) mainly focus on plain network structures. While nodes across networks generally do not have network connections, thus, these network embedding algorithms cannot well capture cross-network proximities and fail to learn generalized representations for prediction tasks across different networks (Heimann et al. 2018). Recently, several attributed network embedding models (Yang, Cohen, and Salakhutdinov 2016; Hamilton, Ying, and Leskovec 2017; Huang, Li, and Hu 2017; Kipf and Welling 2017; Liang et al. 2018) have been proposed to jointly utilize network structures, node attributes and available node labels to learn more informative network representations. Although the attributed network embedding algorithms can capture the proximities between nodes across networks based on node attributes, none of them have considered the domain discrepancy across different networks.

## Cross-network Node Classification

In (Fang, Yin, and Zhu 2013), a network transfer learning algorithm is proposed to project the label propagation matrices of the source network and the target network into a common latent space via Nonnegative Matrix Tri-Factorization technique. In (Shen, Chung, and Mao 2017; Shen, Mao, and Chung 2019), a cross-network learning model is proposed to leverage the useful knowledge learned from a source network to predict seed nodes and inactive edges for influence maximization in a target network. Recently, Shen et al. (2019) proposed a CDNE model to incorporate MMD-based domain adaptation technique into deep network embedding to learn label-discriminative and network-invariant representations for cross-network node classification. In CDNE, different trainable parameters are utilized to learn node representations for the source network and the target network respectively. Most recently, a AdaGCN model (Dai et al. 2019) is proposed to leverage adversarial domain adaptation and graph convolution networks to address cross-network node classification. The proposed ACDNE model is distinct from AdaGCN in terms of both network embedding and domain adaptation. On one hand, for network embedding, AdaGCN employs graph convolution networks (Kipf and Welling 2017; Li et al. 2019) to integrate network topology and node attributes in a semi-supervised learning model. While ACDNE proposes a novel deep network embedding module with two feature extractors to learn latent representations from each node's own attributes and its neighbors' attributes respectively. On the other hand, to reduce domain discrepancy, AdaGCN uses WDGRL (Shen et al. 2018) while ACDNE employs DANN (Ganin et al. 2016).

## Problem Definition

Let $\mathcal{G} = (V, E, A, X, Y)$ denote a network with a set of nodes $V$ and a set of edges $E$. $A \in R^{n \times n}, X \in R^{n \times w}$ and $Y \in R^{n \times c}$ denote the topological proximity matrix, node attribute matrix and node label matrix associated with $\mathcal{G}$, where $n$ is the number of nodes, $w$ is the number of node attributes and $c$ is the number of node labels in $\mathcal{G}$. The $i$-th row of $A, X, Y$, denoted as $a_i, x_i, y_i$, capture the topological proximities, attributes and observable labels associated with node $v_i \in V$. In cross-network node classification problem, we have a fully labeled source network $\mathcal{G}^s = (V^s, E^s, A^s, X^s, Y^s)$ and an unlabeled target network $\mathcal{G}^t = (V^t, E^t, A^t, X^t)$, where the label categories should be the same between two networks. In addition, no common nodes are shared between $\mathcal{G}^s$ and $\mathcal{G}^t$, and no edges are connecting nodes from $\mathcal{G}^s$ and $\mathcal{G}^t$. When two networks do not share the same set of node attributes, one can construct a union attribute set between the attributes from the source network and from the target network. Then, the cross-network embeddings can be learned based on the union attribute set. Note that the data distributions of network connections, node attributes and node labels are generally varied across networks. The goal of *cross-network node classification* is to learn appropriate node representations based upon which the abundant labeled information from the source network can be successfully leveraged to predict node labels for the target network.

## Adversarial Cross-network Deep Network Embedding

Figure 1 shows the architecture of the proposed ACDNE model. It contains three main components, i.e., deep network embedding, node classifier and domain discriminator.

## Deep Network Embedding

The deep network embedding module contains two feature extractors, a concatenation layer and a pairwise constraint.

### Feature Extractors

Firstly, given each node's attributes as input, the first feature extractor (FE1) with $l_f$ hidden layers is constructed as:

$$h_{f_1}^{(k)}(x_i) = ReLU\big(h_{f_1}^{(k-1)}(x_i)W_{f_1}^{(k)} + b_{f_1}^{(k)}\big), 1 \leq k \leq l_f \quad (1)$$

where $h_{f_1}^{(0)}(x_i) = x_i \in R^{1 \times w}$ represents the input attribute vector of $v_i$. $x_{ik}$ is the $k$-th attributed value of $v_i$ and $x_{ik} = 0$ indicates $v_i$ is not associated with the $k$-th attribute. $h_{f_1}^{(k)}(x_i) \in R^{1 \times f(k)}, 1 \leq k \leq l_f$ represents the latent node attribute representation of $v_i$ learned by the $k$-th hidden layer of FE1, and $f(k)$ is the dimensionality of the $k$-th hidden layer of FE1. $W_{f_1}^{(k)}$ and $b_{f_1}^{(k)}$ denote the trainable weight and bias parameters associated with the $k$-th hidden layer of FE1. $ReLU(\cdot)$ is a non-linear activation function characterized by $ReLU(x) = \max(0, x)$.

Secondly, given the neighbors' attributes as the input, the second feature extractor (FE2) with $l_f$ hidden layers is

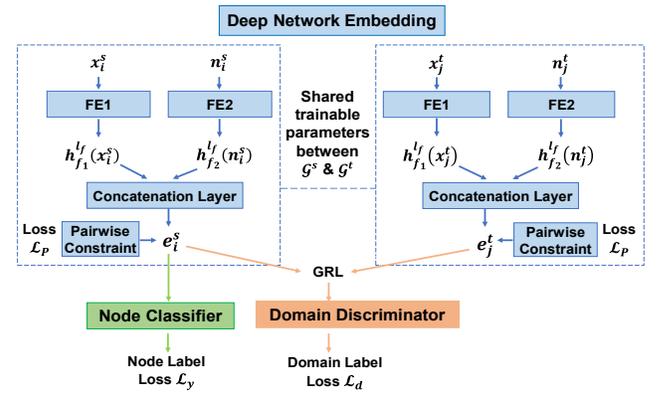

Figure 1. Model architecture of ACDNE. The superscript *s* and *t* denote nodes from the source network and from the target network, respectively.

constructed as:
$$h_{f_2}^{(k)}(n_i) = ReLU\left(h_{f_2}^{(k-1)}(n_i)W_{f_2}^{(k)} + b_{f_2}^{(k)}\right), 1 \le k \le l_f \quad (2)$$
where $h_{f_2}^{(0)}(n_i) = n_i \in R^{1 \times w}$ represents the input neighbor attribute vector of $v_i$. To compute $n_i$, we aggregate the neighbors' attributes by assigning higher weight to closer neighbor (i.e. possessing higher topological proximity with $v_i$), as below:
$$n_{ik} = \sum_{j=1, j \ne i}^{n} \frac{a_{ij}}{\sum_{g=1, g \ne i}^{n} a_{ig}} x_{jk} \quad (3)$$
where $a_{ij}$ denotes the topological proximity between $v_i$ and $v_j$. In this work, we followed (Cao, Lu, and Xu 2016; Shen et al. 2019) to employ the PPMI metric (Levy and Goldberg 2014) to measure the topological proximity between nodes within $K$ steps in a network. A higher positive value of $a_{ij}$ indicates closer network relationship between $v_i$ and $v_j$, while $a_{ij} = 0$ indicates that $v_j$ is not a neighbor of $v_i$ within $K$ steps in network $\mathcal{G}$. $h_{f_2}^{(k)}(n_i) \in R^{1 \times f(k)}$ represents the latent neighbor attribute representation of $v_i$, learned by the $k$-th hidden layer of FE2. $W_{f_2}^{(k)}$ and $b_{f_2}^{(k)}$ denote the trainable parameters associated with the $k$-th hidden layer of FE2. In ACDNE, the number of hidden layers $l_f$ and the dimensionality of each $k$-th hidden layer $f(k), \forall 1 \le k \le l_f$ are set as the same for FE1 and FE2.

**Concatenation Layer**
Next, we feed the deepest latent node attribute representation learned by FE1 i.e. $h_{f_1}^{(l_f)}(x_i)$ and the deepest latent neighbor attribute representation learned by FE2 i.e. $h_{f_2}^{(l_f)}(n_i)$ to a concatenation layer as below:
$$e_i = ReLU\left(\left[h_{f_1}^{(l_f)}(x_i), h_{f_2}^{(l_f)}(n_i)\right]W_c + b_c\right) \quad (4)$$
where $e_i \in R^{1 \times d}$ denotes node representation of $v_i$ finally learned by ACDNE, and $d$ is the embedding dimensionality. $\left[h_{f_1}^{(l_f)}(x_i), h_{f_2}^{(l_f)}(n_i)\right]$ represents the concatenation of $h_{f_1}^{(l_f)}(x_i)$ and $h_{f_2}^{(l_f)}(n_i)$. $W_c, b_c$ are the trainable parameters associated with the concatenation layer. On one hand, by utilizing each node's own attributes as the input in FE1, nodes sharing similar attributes will have similar latent node attribute representations, no matter whether they have network connections or not. On the other hand, by utilizing the neighbors' attributes as the input in FE2, the nodes sharing similar neighborhood or their neighbors sharing similar attributes will have similar latent neighbor attribute representations. Then, by integrating the latent representations learned by FE1 and FE2 to learn final node representations after the concatenation layer, both the attributed affinity and topological proximities between nodes can be well preserved.

**Pairwise Constraint**
Next, we incorporate the following pairwise constraint on node representations to explicitly preserve the topological proximities between nodes within each network:
$$\mathcal{L}_p = \frac{1}{n^s}\sum_{v_i, v_j \in V^s} a_{ij}\|e_i - e_j\|^2 + \frac{1}{n^t}\sum_{v_i, v_j \in V^t} a_{ij}\|e_i - e_j\|^2 \quad (5)$$
where $n^s$ and $n^t$ denote the number of nodes in $\mathcal{G}^s$ and $\mathcal{G}^t$, respectively. By minimizing $\mathcal{L}_p$, more strongly connected nodes within the source network or within the target network would have more similar node representations. For simplicity, we denote all the trainable parameters associated with the aforementioned deep network embedding module as $\theta_e = \left\{\left\{W_{f_1}^{(k)}, b_{f_1}^{(k)}, W_{f_2}^{(k)}, b_{f_2}^{(k)}\right\}_{k=1}^{l_f}, W_c, b_c\right\}$.

**Node Classifier**
To make node representations label-discriminative, we incorporate the supervised signals from the source network, by adding a node classifier at the top of the deep network embedding module, as:
$$\hat{y}_i = \phi(e_i W_y + b_y) \quad (6)$$
where $\hat{y}_i \in R^{1 \times c}$ denotes the predicted probabilities of $v_i$ over the $c$ label categories. $\phi(\cdot)$ is the output function of the classifier, one can employ Softmax function for multi-class classification or Sigmoid function for multi-label classification. $\theta_y = \{W_y, b_y\}$ represents the trainable parameters associated with node classification. By utilizing all the labeled nodes from the source network for training, the Softmax cross-entropy loss is defined for multi-class node classification, as:
$$\mathcal{L}_y = -\frac{1}{n^s}\sum_{v_i \in V^s}\sum_{k=1}^{c} y_{ik} log(\hat{y}_{ik}) \quad (7)$$
where $y_{ik}$ denotes the ground-truth label of $v_i$, $y_{ik} = 1$ if $v_i$ is associated with label $k$; otherwise, $y_{ik} = 0$. $\hat{y}_{ik}$ represents the predicted probability of $v_i$ to be labeled with category $k$. In addition, for multi-label node classification, the one-vs-rest Sigmoid cross-entropy loss is defined as:
$$\mathcal{L}_y = -\frac{1}{n^s}\sum_{v_i \in V^s}\sum_{k=1}^{c} y_{ik} log(\hat{y}_{ik}) + (1 - y_{ik})log(1 - \hat{y}_{ik}) \quad (8)$$

**Adversarial Domain Adaptation**
Next, we employ an adversarial domain adaptation approach to make node representations learned by ACDNE network-invariant. Firstly, one can feed the node representation learned by the deep network embedding module to a domain discriminator to predict which network a node comes from, as follows:
$$h_d^{(k)}(e_i) = ReLU\left(h_d^{(k-1)}(e_i)W_d^{(k)} + b_d^{(k)}\right), 1 \le k \le l_d$$
$$\hat{d}_i = Softmax\left(h_d^{(l_d)}(e_i)W_d^{(l_d+1)} + b_d^{(l_d+1)}\right) \quad (9)$$
where $h_d^{(0)}(e_i) = e_i$, $h_d^{(k)}(e_i) \in R^{1 \times d(k)}$ represents the domain representation of $v_i$ learned by the $k$-th hidden layer of the domain discriminator, $d(k)$ is the dimensionality of the $k$-th hidden layer, and $l_d$ is the number of hidden layers in the domain discriminator. $\theta_d = \{W_d^{(k)}, b_d^{(k)}\}_{k=1}^{l_d+1}$ represents the trainable parameters associated with the domain discriminator. Then, by utilizing nodes from the source network as well as from the target network for training, the domain classification loss is defined as:

$$\mathcal{L}_d = -\frac{1}{n^s+n^t}\sum_{v_i \in \{V^s \cup V^t\}}(1-d_i)log(1-\hat{d}_i) + d_i log(\hat{d}_i) \quad (10)$$

where $d_i$ is the ground-truth domain label of $v_i$, $d_i = 1$ if $v_i \in V^t$ and $d_i = 0$ if $v_i \in V^s$. $\hat{d}_i$ represents the predicted probability of $v_i$ coming from the target network. To make node representations network-invariant, the domain discriminator and deep network embedding module compete against each other in an adversarial manner. On one hand, $\min_{\theta_d}\{\mathcal{L}_d\}$ enables the domain discriminator to accurately distinguish the node representations of the source network from those of the target network. On the other hand, $\min_{\theta_e}\{-\mathcal{L}_d\}$ makes the deep network embedding module trained to deceive the domain discriminator by generating node representations which are indistinguishable across networks.

---

**Algorithm 1: ACDNE**

**Input**: Fully labeled source network $\mathcal{G}^s = (V^s, E^s, A^s, X^s, Y^s)$ and unlabeled target network $\mathcal{G}^t = (V^t, E^t, A^t, X^t)$; batch size $b$, pairwise constraint weight $p$, domain adaptation weight $\lambda$.

1     while not max iteration do:
2        for each mini-batch $B$ do:
3           for $v_i^s \in V^s$ in $B$:
4              Learn node attribute representation by FE1 and neighbor attribute representation by FE2;
5              Learn node representation by concatenation layer, $e_i^s$;
6           end for
7           for $v_j^t \in V^t$ in $B$:
8              Learn node attribute representation by FE1 and neighbor attribute representation by FE2;
9              Learn node representation by concatenation layer, $e_j^t$;
10          end for
11          Compute pairwise constraint loss $\mathcal{L}_p$ based on $\{(e_i^s, a_i^s)\}_{i=1}^{b/2}$ and $\{(e_j^t, a_j^t)\}_{j=1}^{b/2}$;
12          Compute node classification loss $\mathcal{L}_y$ based on $\{(e_i^s, y_i^s)\}_{i=1}^{b/2}$;
13          Compute domain classification loss $\mathcal{L}_d$ based on $\{(e_i^s, d_i)\}_{i=1}^{b/2}$ and $\{(e_j^t, d_j)\}_{j=1}^{b/2}$;
14          Update parameters $\theta_e, \theta_y, \theta_d$ via SGD in (12);
15        end for
16    end while

**Output**: Optimized parameters $\theta_e^*, \theta_y^*, \theta_d^*$; Cross-network node representations $\{e_i^s\}_{i=1}^{n^s}$ and $\{e_j^t\}_{j=1}^{n^t}$; Predicted node labels for $\mathcal{G}^t$: $\{\hat{y}_j^t\}_{j=1}^{n^t}$.

---

## Joint Training

By integrating deep network embedding, node classifier and adversarial domain adaptation, the goal of ACDNE is to optimize the following minimax objective:

$$\min_{\theta_e, \theta_y}\left\{\mathcal{L}_y + p\mathcal{L}_p + \lambda \max_{\theta_d}\{-\mathcal{L}_d\}\right\} \quad (11)$$

where $p, \lambda$ are the trade-off parameters to balance the effects of different terms. In this work, we follow (Ganin et al. 2016) to insert a Gradient Reversal Layer (GRL) between the deep network embedding module and the domain discriminator so as to simultaneously update them during backpropagation. The GRL reverses the partial derivative of the domain classification loss $\mathcal{L}_d$ w.r.t. the network embedding parameters $\theta_e$ and multiplies them by a coefficient $\lambda$. Then, ACDNE can be optimized by stochastic gradient descent (SGD) as follows:

$$\theta_e \leftarrow \theta_e - \mu\left(\frac{\partial \mathcal{L}_y}{\partial \theta_e} + p\frac{\partial \mathcal{L}_p}{\partial \theta_e} - \lambda\frac{\partial \mathcal{L}_d}{\partial \theta_e}\right)$$
$$\theta_y \leftarrow \theta_y - \mu\frac{\partial \mathcal{L}_y}{\partial \theta_y} \quad (12)$$
$$\theta_d \leftarrow \theta_d - \mu\frac{\partial \mathcal{L}_d}{\partial \theta_d}$$

where $\mu$ denotes the learning rate. Algorithm 1 shows the training process of ACDNE. Firstly, in each mini-batch, we sample half nodes from the source network and half nodes from the target network. Then, the same deep network embedding module is employed to learn node representations for two networks, in Lines 3-10. Then, the pairwise constraint loss, node classification loss and domain classification loss are computed for each mini-batch in Lines 11-13. Next, the trainable parameters of ACDNE are updated by SGD in Line 14. After ACDNE finally converges or a maximum training iteration has been reached, one can employ the optimized network embedding parameters $\theta_e^*$ to generate label-discriminative and network-invariant node representations across networks, i.e., $\{e_i^s\}_{i=1}^{n^s}$ and $\{e_j^t\}_{j=1}^{n^t}$. Next, the optimized node classification parameters $\theta_y^*$ would be employed on $\{e_j^t\}_{j=1}^{n^t}$ to predict node labels for the target network.

## Experiments

### Datasets

ACDNE was evaluated on the cross-network datasets (Shen et al. 2019), the statistics are shown in Table 1. Blog1 and Blog2 are two disjoint social networks extracted from the BlogCatalog dataset (Li et al. 2015), where each node represents a blogger and each edge indicates the friendship between two bloggers. The attributes of each node are the keywords extracted from the blogger's self-description. A node is associated with one label indicating its joining group. Since Blog1 and Blog2 were extracted from the same original network, their data distributions are almost the same. To enlarge domain discrepancy, in each network, 30% of non-zero attributed values were randomly selected to alter as "0" and 30% of zero attributed values were randomly selected to alter as "1" to simulate the incomplete and noisy attributed information across networks. To reduce noises, in the experiments, we employed PCA (Mackiewicz and Ratajczak 1993) as a pre-processing step to extract 1000-D attributes from the original node attributes and employed them as the input attributes for Blog1 and Blog2.

Table 1: Statistics of the networked datasets.

| Dataset | #Nodes | #Edges | #Attributes | #Union Attributes | #Labels |
|---|---|---|---|---|---|
| Blog1 | 2300 | 33471 | 8189 | 8189 | 6 |
| Blog2 | 2896 | 53836 | 8189 | | |
| Citationv1 | 8935 | 15113 | 5379 | 6775 | 5 |
| DBLPv7 | 5484 | 8130 | 4412 | | |
| ACMv9 | 9360 | 15602 | 5571 | | |

On the other hand, Citationv1, DBLPv7 and ACMv9 are three citation networks extracted from the ArnetMiner datasets (Tang et al. 2008), where a node represents a paper and an edge indicates the citation relationship. We modeled the citation networks as undirected networks. The attributes of each node are the sparse bag-of-words features extracted from the paper title. A node can have multiple labels to indicate its relevant research areas. Since Citationv1, DBLPv7 and ACMv9 were extracted from different sources and also formed in different time periods, they inherently have varied data distributions.

In the experiments, two cross-network node classification tasks are performed between Blog1 and Blog2, and six cross-network node classification tasks are conducted among Citationv1, DBLPv7 and ACMv9.

### Baselines

The proposed ACDNE model was benchmarked against the following state-of-the-art algorithms:

Domain Adaptation: **MMD** (Gretton et al. 2007) is incorporated into deep neural network to match the mean of distributions between two domains. **DANN** (Ganin et al. 2016) is an adversarial domain adaptation algorithm which inserts a GRL between the feature extractor and domain discriminator to optimize the minimax objective.

Attributed Network Embedding: **ANRL** (Zhang et al. 2018) learns node representations via a neighbor enhancement autoencoder and an attribute-aware skip-gram model. **LANE** (Huang, Li, and Hu 2017) projects network structures, node attributes and node labels into a unified embedding space via eigenvector decomposition. **SEANO** (Liang et al. 2018) utilizes each node's attributes and the average attributes of its neighborhoods to jointly predict node labels and two types of node contexts localized by graph structure and node label. **GCN** (Kipf and Welling 2017) employs a graph convolutional neural network to utilize network structures, node attributes and node labels to learn node representations.

Cross-network Node Classification: **NetTr** (Fang, Yin, and Zhu 2013) learns the shared structural features across networks by projecting the label propagation matrices of the source network and the target network into a common latent space. **CDNE** (Shen et al. 2019) integrates deep network embedding with MMD-based domain adaptation to learn node representations for the source network and the target network by two stacked auto-encoders respectively.

### Implementation Details

In the experiments, we set $K$-step as 3 when measuring the PPMI topological proximities between nodes within each network. In ACDNE, both FE1 and FE2 are constructed with two hidden layers, with the hidden dimensionalities set as $f(1) = 512, f(2) = 128$. The dimensionality of node representations learned by ACDNE is set as $d = 128$. For fair comparison, the same dimensionality is also set for other baselines. In addition, the domain discriminator is constructed with two hidden layers with dimensionalities as $d(1) = d(2) = 128$. The weight of pairwise constraint $p$ is set as 0.1 for the sparse citation networks and as $10^{-3}$ for the dense Blog networks. Besides, a $L$2-norm regularization term with a weight of $10^{-3}$ is imposed on the trainable weights to prevent overfitting. ACDNE is trained by SGD with a momentum rate of 0.9 over shuffled mini-batches with a batch size of 100. Following (Ganin et al. 2016), the learning rate is decayed as $\mu_p = \frac{\mu_0}{(1+10p)^{0.75}}$, where $\mu_0$ is the initial learning rate (set as 0.01 for the Blog networks and 0.02 for the citation networks), $p$ is the training progress linearly changing from 0 to 1, and the domain adaptation weight $\lambda$ is progressively increased as $\frac{2}{1+\exp(-10p)} - 1$.

### Cross-network Node Classification

In cross-network node classification, a classifier is firstly trained based on all the labeled nodes from the source network, and then tested on all the unlabeled nodes in the target network. Micro-F1 and Macro-F1 are employed as two metrics to evaluate the cross-network node classification performance. In the experiments, each comparing algorithm has been repeatedly run 5 times, and the averaged F1 scores are reported in Table 2.

Firstly, we can see that MMD and DANN achieve the lowest F1 scores in most cross-network node classification tasks. This is because although MMD and DANN can learn domain-invariant representations based on node attributes, during representation learning, they just consider each data sample independently. While in network structural data, considering the complex network relationships between nodes should be rather important and effective for graph mining. Thus, the existing domain adaptation algorithms developed for CV or NLP cannot be directly applied to effectively tackle cross-network node classification. On the other hand, the attributed network embedding algorithms which take full advantage of network topological structures and node attributes can significantly outperform MMD and DANN. In addition, among the attributed network embedding algorithms, GCN achieves the best overall performance. However, GCN still performs much worse than CDNE. This is because GCN does not address domain discrepancy, while CDNE incorporates the MMD-based domain adaptation technique into deep network embedding to reduce the distribution discrepancy across networks. This reflects that in order to achieve good performance in cross-network node classification, both network embedding and domain adaptation are indispensable.

Table 2: Micro-F1 and Macro-F1 scores of cross-network node classification when the source network is fully labeled and the target network is totally unlabeled. The highest F1 scores among all comparing algorithms are shown in Boldface.

| $\mathcal{G}^s$ | $\mathcal{G}^t$ | F1 | MMD | DANN | ANRL | LANE | SEANO | GCN | NetTr | CDNE | ACDNE |
|---|---|---|---|---|---|---|---|---|---|---|---|
| Blog1 | Blog2 | Micro | 0.4385 | 0.4495 | 0.4776 | 0.4703 | 0.4987 | 0.5114 | 0.5014 | **0.6660** | 0.6625 |
| | | Macro | 0.4370 | 0.4484 | 0.4591 | 0.4575 | 0.4959 | 0.4788 | 0.4918 | **0.6643** | 0.6600 |
| Blog2 | Blog1 | Micro | 0.4595 | 0.4656 | 0.4417 | 0.4957 | 0.5023 | 0.4983 | 0.5243 | **0.6384** | 0.6354 |
| | | Macro | 0.4580 | 0.4642 | 0.4226 | 0.4943 | 0.4985 | 0.4634 | 0.5151 | **0.6366** | 0.6351 |
| Citationv1 | DBLPv7 | Micro | 0.5701 | 0.5785 | 0.6603 | 0.5857 | 0.6931 | 0.7124 | 0.5988 | 0.7415 | **0.7735** |
| | | Macro | 0.5358 | 0.5515 | 0.6278 | 0.5508 | 0.6694 | 0.6812 | 0.5518 | 0.7171 | **0.7609** |
| DBLPv7 | Citationv1 | Micro | 0.5340 | 0.5627 | 0.6664 | 0.5695 | 0.7150 | 0.7163 | 0.5911 | 0.7961 | **0.8209** |
| | | Macro | 0.4962 | 0.5413 | 0.6344 | 0.5383 | 0.6954 | 0.6719 | 0.5553 | 0.7805 | **0.8025** |
| Citationv1 | ACMv9 | Micro | 0.5416 | 0.5553 | 0.6446 | 0.5627 | 0.6781 | 0.7132 | 0.5775 | 0.7752 | **0.7956** |
| | | Macro | 0.5115 | 0.5345 | 0.6202 | 0.5300 | 0.6625 | 0.6919 | 0.5344 | 0.7679 | **0.7888** |
| ACMv9 | Citationv1 | Micro | 0.5448 | 0.5673 | 0.6841 | 0.5802 | 0.7203 | 0.7356 | 0.5881 | 0.7891 | **0.8327** |
| | | Macro | 0.5201 | 0.5492 | 0.6577 | 0.5517 | 0.7029 | 0.7003 | 0.5546 | 0.7700 | **0.8166** |
| DBLPv7 | ACMv9 | Micro | 0.5143 | 0.5311 | 0.6308 | 0.5362 | 0.6664 | 0.6683 | 0.5623 | **0.7659** | 0.7634 |
| | | Macro | 0.4651 | 0.5007 | 0.6019 | 0.4924 | 0.6528 | 0.6291 | 0.5099 | 0.7591 | **0.7609** |
| ACMv9 | DBLPv7 | Micro | 0.5448 | 0.5535 | 0.6448 | 0.5706 | 0.6613 | 0.6822 | 0.5630 | 0.7203 | **0.7657** |
| | | Macro | 0.5116 | 0.5249 | 0.6103 | 0.5256 | 0.6333 | 0.6413 | 0.4980 | 0.6978 | **0.7431** |

In addition, the inputs of SEANO (i.e. attributes of each node and its neighborhood) are similar to that of the deep network embedding module in ACDNE. However, ACDNE outperforms SEANO by a large margin. This is because unlike SEANO, ACDNE further incorporates a concatenation layer and pairwise constraint into deep network embedding module to learn more informative representations. In addition, besides node classification, SEANO also predicts each node's neighborhood as one of the outputs. This architecture makes SEANO focus more on preserving the proximities between nodes naturally having network connections, while nodes across networks do not have network connections. Thus, SEANO would have limited ability to capture cross-network proximities. Furthermore, in contrast to SEANO, ACDNE also employs an adversarial domain adaptation approach to learn network-invariant representations. The significant outperformance of ACDNE over SEANO again verifies the necessity of reducing domain discrepancy in cross-network node classification.

Next, we discuss the performance of the algorithms developed for cross-network node classification. As shown in Table 2, NetTr achieves much worse performance than both CDNE and ACDNE. This is because NetTr learns the common latent features across networks based on topological structures only, while the same labeled nodes from different networks can have very distinct topological structures. In addition, one can see that the proposed ACDNE model achieves comparable performance w.r.t. CDNE in three tasks and significantly outperforms CDNE in the other tasks. This is because unlike CDNE which utilizes MMD to minimize domain discrepancy, ACDNE employs a more powerful adversarial domain adaptation approach. The outperformance of DANN over MMD also verifies this. In addition, CDNE utilizes topological structures to capture within-network proximities, while only leverages the node labels predicted based on node attributes to capture cross-network proximities. If the node attributes are rather noisy or incomplete, then the predicted node labels would be inaccurate, which would further yield inaccurate cross-network alignment. In contrast, in ACDNE, two feature extractors are utilized to learn representations based on each node's attributes and its neighbors' attributes. Such architecture can effectively alleviate the negative effects caused by the noisy or incomplete attributed information.

Table 3: Micro-F1 score of ACDNE variants.

| Model Variant | Blog1→Blog2 | Citationv1→ACMv9 |
|---|---|---|
| ACDNE | 0.6625 | 0.7956 |
| Without FE1 | 0.5021 | 0.7791 |
| Without FE2 | 0.4434 | 0.6210 |
| Without pairwise constraint | 0.6347 | 0.7677 |
| Without node classifier | 0.2617 | 0.4307 |
| Without discriminator | 0.5402 | 0.7481 |

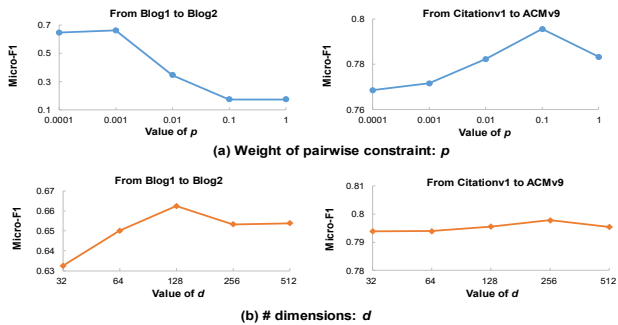

Figure 2. Sensitivities of parameters $p$ and $d$ on the performance of ACDNE.

## Ablation Test

Next, we conduct ablation studies to investigate the effectiveness of different components in ACDNE. As shown in Table 3, without either FE1 or FE2, the Micro-F1 scores would be significantly dropped as compared to ACDNE. This demonstrates the effectiveness of employing two feature extractors in the proposed deep network embedding module. Also, the worse performance of without pairwise constraint as compared to ACDNE reflects that explicitly preserving the topological proximities between nodes within each network can effectively yield informative representations for node classification. Moreover, without node classifier performs significantly worse than ACDNE. This demonstrates that incorporating the supervised signals from the source network to learn label-discriminative representations is indeed essential for cross-network node classification. Lastly, without domain discriminator yields much worse performance than ACDNE. This demonstrates that reducing domain discrepancy is indeed necessary for cross-network node classification.

## Parameter Sensitivity

Parameter $p$ denotes the weight of pairwise constraint to preserve topological proximities between nodes within each network. As shown in Figure 2(a), the performance of ACDNE is sensitive to the value of $p$. It is suggested to set relatively small value of $p$ (i.e. $10^{-3}$) for the dense Blog networks, while set relatively large value of $p$ (i.e. $10^{-1}$) for the sparse citation networks. Parameter $d$ denotes the dimensionality of node representations learned by ACDNE. As shown in Figure 2(b), in the Blog networks, setting $d=128$ yields better performance than other values. While in the citation networks, different values of $d$ in $\{32, 64, 128, 256, 512\}$ can all achieve good performance for ACDNE.

## Visualization

Next, we employ the t-SNE toolkit (Maaten and Hinton 2008) to visualize node representations learned by ACDNE

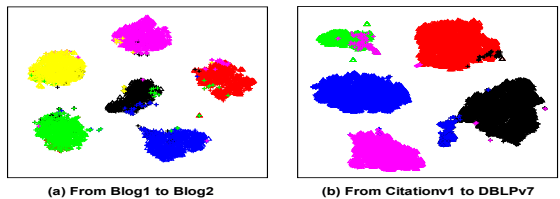

Figure 3. Visualization of node representations learned by AC-DNE. Different colors correspond to different node labels. The triangle symbols represent nodes from the source network while the plus symbols correspond to nodes from the target network.

in a 2-D space. As shown in Figure 3, nodes belonging to different categories have been mostly mapped into separable areas. This indicates that the node representations learned by ACDNE are indeed label-discriminative. On the other hand, the same labeled nodes across networks have been mostly mapped to the same area. This reflects that the node representations learned by ACDNE are actually network-invariant. Besides, a few different colored nodes are also mapped close. This is because different categories of nodes are also possible to have network connections or share similar attributes.

## Conclusions

In this paper, we propose a novel ACDNE model to integrate deep network embedding with the emerging adversarial domain adaptation technique to address cross-network node classification. The deep network embedding module employs two feature extractors to 1) make nodes sharing similar attributes have similar latent node attribute representations independent of their network positions; and 2) make nodes sharing similar neighborhood or their neighborhood sharing similar attributes have similar latent neighbor attribute representations. As a result, both attributed affinity and topological proximities between nodes can be well preserved. A node classifier is incorporated to leverage the supervised signals from the source network to guide the node representations learned by ACDNE to be label-discriminative. In addition, a domain discriminator is incorporated into ACDNE to compete against the deep network embedding module to make node representations network-invariant. The extensive experimental results in the real-world datasets demonstrate the distinctive performance of ACDNE over the state-of-the-art algorithms.

## Acknowledgments

This work was supported in part by Innovative Technology Fund (No. MRP/015/18), UGC PolyU (No. 152071/17E), National Natural Science Foundation of China (No. 71901050) and the General Programs of the Sichuan Province Social Science Association (No. SC19B035).

# References

Cao, S.; Lu, W.; and Xu, Q. 2016. Deep Neural Networks for Learning Graph Representations. In *AAAI Conference on Artificial Intelligence*, 1145-1152.

Dai, Q.; Li, Q.; Tang, J.; and Wang, D. 2018. Adversarial Network Embedding. In *AAAI Conference on Artificial Intelligence*.

Dai, Q.; Shen, X.; Wu, X.-M.; and Wang, D. 2019. Network Transfer Learning via Adversarial Domain Adaptation with Graph Convolution. *arXiv preprint arXiv:1909.01541*.

Dai, Q.; Shen, X.; Zhang, L.; Li, Q.; and Wang, D. 2019. Adversarial Training Methods for Network Embedding. In *International Conference on World Wide Web*.

Dai, W.; Yang, Q.; Xue, G.-R.; and Yu, Y. 2007. Boosting for Transfer Learning. In *International Conference on Machine Learning*, 193-200.

Fang, M.; Yin, J.; and Zhu, X. 2013. Transfer Learning across Networks for Collective Classification. In *International Conference on Data Mining*, 161-170.

Ganin, Y.; Ustinova, E.; Ajakan, H.; Germain, P.; Larochelle, H.; Laviolette, F.; Marchand, M.; and Lempitsky, V. 2016. Domain-adversarial Training of Neural Networks. *The Journal of Machine Learning Research* 17(1): 2096-2030.

Goodfellow, I.; Pouget-Abadie, J.; Mirza, M.; Xu, B.; Warde-Farley, D.; Ozair, S.; Courville, A.; and Bengio, Y. 2014. Generative Adversarial Nets. In *Advances in Neural Information Processing Systems*, 2672-2680.

Gretton, A.; Borgwardt, K. M.; Rasch, M.; Schölkopf, B.; and Smola, A. J. 2007. A Kernel Method for the Two-Sample-Problem. In *Advances in Neural Information Processing Systems*, 513-520.

Grover, A., and Leskovec, J. 2016. node2vec: Scalable Feature Learning for Networks. In *ACM SIGKDD International Conference on Knowledge Discovery and Data Mining*, 855-864.

Hamilton, W.; Ying, Z.; and Leskovec, J. 2017. Inductive Representation Learning on Large Graphs. In *Advances in Neural Information Processing Systems*, 1024-1034.

Heimann, M.; Shen, H.; Safavi, T.; and Koutra, D. 2018. REGAL: Representation Learning-based Graph Alignment. In *ACM Conference on Information and Knowledge Management*.

Huang, J.; Gretton, A.; Borgwardt, K.; Schölkopf, B.; and Smola, A. J. 2007. Correcting Sample Selection Bias by Unlabeled Data. In *Advances in Neural Information Processing Systems*, 601-608.

Huang, X.; Li, J.; and Hu, X. 2017. Label Informed Attributed Network Embedding. In *ACM International Conference on Web Search and Data Mining*, 731-739.

Kipf, T. N., and Welling, M. 2017. Semi-supervised Classification with Graph Convolutional Networks. In *International Conference on Learning Representations*.

Levy, O., and Goldberg, Y. 2014. Neural Word Embedding as Implicit Matrix Factorization. In *Advances in Neural Information Processing Systems*, 2177-2185.

Li, J.; Hu, X.; Tang, J.; and Liu, H. 2015. Unsupervised Streaming Feature Selection in Social Media. In *ACM Conference on Information and Knowledge Management*, 1041-1050.

Li, Q.; Wu, X.-M.; Liu, H.; Zhang, X.; and Guan, Z. 2019. Label Efficient Semi-Supervised Learning via Graph Filtering. In *IEEE Conference on Computer Vision and Pattern Recognition*.

Liang, J.; Jacobs, P.; Sun, J.; and Parthasarathy, S. 2018. Semi-supervised Embedding in Attributed Networks with Outliers. In *SIAM International Conference on Data Mining*, 153-161.

Long, M.; Cao, Y.; Wang, J.; and Jordan, M. I. 2015. Learning Transferable Features with Deep Adaptation Networks. In *International Conference on Machine Learning*, 97-105.

Long, M.; Wang, J.; Ding, G.; Sun, J.; and Yu, P. S. 2013. Transfer Feature Learning with Joint Distribution Adaptation. In *IEEE International Conference on Computer Vision*, 2200-2207.

Maaten, L. v. d., and Hinton, G. 2008. Visualizing Data Using t-SNE. *Journal of Machine Learning Research* 9(Nov): 2579-2605.

Mackiewicz, A., and Ratajczak, W. 1993. Principal Components Analysis (PCA). *Computers and Geosciences* 19: 303-342.

Pan, S. J., and Yang, Q. 2010. A Survey on Transfer Learning. *IEEE Transactions on Knowledge and Data Engineering* 22(10): 1345-1359.

Perozzi, B.; Al-Rfou, R.; and Skiena, S. 2014. Deepwalk: Online Learning of Social Representations. In *ACM SIGKDD International Conference on Knowledge Discovery and Data Mining*, 701-710.

Shen, J.; Qu, Y.; Zhang, W.; and Yu, Y. 2018. Wasserstein Distance Guided Representation Learning for Domain Adaptation. In *AAAI Conference on Artificial Intelligence*.

Shen, X., and Chung, F.-l. 2017. Deep Network Embedding with Aggregated Proximity Preserving. In *IEEE/ACM International Conference on Advances in Social Network Analysis and Mining*, 40-44.

Shen, X., and Chung, F.-L. 2018. Deep Network Embedding for Graph Representation Learning in Signed Networks. *IEEE Transactions on Cybernetics*. DOI: 10.1109/TCYB.2018.2871503.

Shen, X.; Chung, F.-l.; and Mao, S. 2017. Leveraging Cross-Network Information for Graph Sparsification in Influence Maximization. In *ACM SIGIR Conference on Research and Development in Information Retrieval*, 801-804.

Shen, X.; Dai, Q.; Mao, S.; Chung, F.-l.; and Choi, K.-S. 2019. Network Together: Node Classification via Cross network Deep Network Embedding. *arXiv preprint arXiv:1901.07264*.

Shen, X.; Mao, S.; and Chung, F.-l. 2019. Cross-network Learning with Fuzzy Labels for Seed Selection and Graph Sparsification in Influence Maximization. *IEEE Transactions on Fuzzy Systems*. DOI: 10.1109/TFUZZ.2019.2931272.

Tang, J.; Qu, M.; Wang, M.; Zhang, M.; Yan, J.; and Mei, Q. 2015. Line: Large-scale Information Network Embedding. In *International Conference on World Wide Web*, 1067-1077.

Tang, J.; Zhang, J.; Yao, L.; Li, J.; Zhang, L.; and Su, Z. 2008. Arnetminer: Extraction and Mining of Academic Social Networks. In *ACM SIGKDD International Conference on Knowledge Discovery and Data Mining*, 990-998.

Tzeng, E.; Hoffman, J.; Saenko, K.; and Darrell, T. 2017. Adversarial Discriminative Domain Adaptation. In *IEEE Conference on Computer Vision and Pattern Recognition*, 7167-7176.

Wang, D.; Cui, P.; and Zhu, W. 2016. Structural Deep Network Embedding. In *ACM SIGKDD International Conference on Knowledge Discovery and Data Mining*, 1225-1234.

Yang, Z.; Cohen, W. W.; and Salakhutdinov, R. 2016. Revisiting Semi-supervised Learning with Graph Embeddings. *International Conference on Machine Learning*.

Zhang, Z.; Yang, H.; Bu, J.; Zhou, S.; Yu, P.; Zhang, J.; Ester, M.; and Wang, C. 2018. ANRL: Attributed Network Representation Learning via Deep Neural Networks. In *International Joint Conference on Artificial Intelligence*, 3155-3161.